# Generalized rainbow patterns of oblate drops simulated by a ray model in three dimensions

QINGWEI DUAN[1,2], FABRICE R.A. ONOFRI[3], XIANG'E HAN[1], AND KUAN FANG REN[2,*]

[1]School of Physics and Optoelectronic Engineering, Xidian University, Xi'an 710071, China
[2]CORIA-UMR 6614, Normandie Université, CNRS, Université et INSA de Rouen, 76801 Saint-Etienne du Rouvray, France
[3]Aix-Marseille Université, CNRS, IUSTI, UMR 7343, 13453 Marseille, France
[*]Corresponding author: fang.ren@coria.fr

Compiled June 7, 2021

**The scattering patterns near the primary rainbow of oblate drops are simulated by extending the vectorial complex ray model (VCRM) [1] to three-dimensional (3D) calculations. With the curvature of wavefront as intrinsic property of a ray, this advanced ray model permits, in principle, to predict the amplitudes and phases of all emergent rays with a rigorous algebraic formalism. This letter reports a breakthrough of VCRM for 3D scattering with a line-by-line triangulation interpolation algorithm allowing to calculate the total complex amplitude of scattered field. This makes possible to simulate not only the skeleton (geometrical rainbow angles, hyperbolic-umbilic caustics), but also the coarse (Airy bows, lattice) and fine (ripple fringes) structures of the generalized rainbow patterns (GRPs) of oblate drops. The simulated results are found qualitatively and quantitatively in good agreement with experimental scattering patterns for drops of different aspect ratios. The physical interpretation of the GRPs is also given. This work opens up prominent perspectives for simulating and understanding the 3D scattering of large particles of any shape with smooth surface by VCRM.**

The scattering patterns near the primary rainbow of water drops of oblate spheroidal shape were revealed experimentally by Marston et al. [2]. Some of the patterns, like the hyperbolic-umbilic (HU) diffraction catastrophe and the hexagonal lattice [3], are not observed in the rainbows of spherical drops and hence referred to as the GRPs. Their prediction and interpretation are of theoretical importance since similar patterns are frequently observed in the scattering of non-spherical objects. They have also significant impact on the optical particle characterization [4, 5]. From pure geometrical optics (GO), Nye [6] derived relations to predict the caustic structures of the GRPs. Later, Yu et al. [4, 7] developed a vector ray tracing model allowing to predict the caustic structures of GRPs numerically according to the limiting angles of a large number of rays. However, this ray model and the pure GO method fail to calculate the amplitudes and the phases of the emergent rays. On the other hand, there exist electromagnetic solutions for the scattering of spheroids, such as that based on Debye series expansion [8]. Unfortunately, because of numerical difficulty, this approach is limited to small and moderately elongated spheroids (equivalent-volume sphere radius less than a few dozens of wavelengths). Since droplets in this size range are usually perfectly spherical due to the surface tension, it is clear that this approach falls to meet the present demand (drops from dozens to thousands of wavelengths). Finally, physical optics or diffraction model [3, 9–11] may be used to calculate the scattering patterns in the vicinity of caustics, but they are applicable to a limited region and require complex analytical derivations, so not generalizable to arbitrarily shaped and oriented particles. These are some of the reasons why the development of ray models is still necessary.

The main difficulties in using ray models [4, 7, 12–14] to predict the scattering of a non-spherical particle are to calculate the divergence factor and to determine the phase shift due to focal lines for each emergent ray. To address these problems, Ren et al. [1] proposed VCRM. In this advanced model, the curvature of wavefront is considered as an intrinsic physical property of a light ray, and all properties of a ray are expressed in a vectorial and algebraic formalism, simplifying and speeding up the calculation for non-spherical particles. Though its formalism is for 3D scattering, the numerical implementation was limited to the scattering in a symmetric plane, namely in the two cases: the scattering of infinite cylinders at normal incidence [15] or the scattering of ellipsoids in the equatorial plane [1]. In the latter case, the curvature in the direction perpendicular to the symmetric plane is taken into account but those rays out of the symmetric plane are not considered. For clarity, we refer in this letter the former as VCRM2D and the latter as VCRM2D+. The predictions of VCRM2D+ were used with success to predict the scattering diagrams in the inter-caustic zone of oblate drops trapped in an acoustic field [5], allowing the determination of the aspect ratio and refractive index of the drop. The present letter reports the extension of VCRM to 3D scattering (VCRM3D), thereby providing a feasible way to predict and understand the 3D scattering patterns of large non-spherical particles. The case of large oblate drops is taken for comparison purpose and as a generic case for non-spherical particles.

The surface of an oblate drop is expressed by the canonical equation $(x^2 + y^2)/a^2 + z^2/c^2 = 1$, where a and c are respectively the semi-axes in the horizontal equatorial plane and along the vertical symmetry axis. The drop in air with refractive index m is illuminated by a horizontally propagating plane wave of wavelength $\lambda$. The direction of the incident wave is set as the x axis of Cartesian coordinate system. In the ray model, the incident plane wave is simulated by bundles of rays which are regularly distributed. The direction of an emergent ray in 3D space is described [7] by the azimuth angle $\theta$ and the elevation angle $\psi$, as illustrated in Fig. 1. The order of an emergent ray is noted by p with p = 0 for the externally reflected rays and p ≥ 1 for the emergent rays undergone p − 1 internal reflections [11].

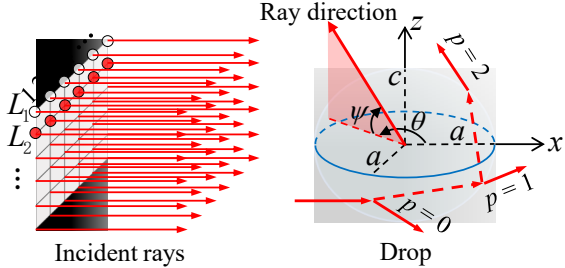

**Fig. 1.** Schema of light scattering by an oblate drop using ray model.

In VCRM, the electric vector of an emergent ray can be expressed as

$$\begin{pmatrix} E_\perp^e \exp(i\Phi_\perp^e) \\ E_k^e \exp(i\Phi_k^e) \end{pmatrix} = k\sqrt{D} \begin{pmatrix} E_\perp \exp(i\Phi_\perp) \\ E_k \exp(i\Phi_k) \end{pmatrix} \exp[i(\Phi_P + \Phi_F)] \quad (1)$$

where the subscripts $\perp$ and k stand for the components perpendicular and parallel to the scattering plane respectively. $k = 2\pi/\lambda$ is the wave number. The divergence factor D, which counts the intensity variation due to the divergence or convergence of wave, is determined in VCRM by the Gaussian curvature of the wavefront [1, 16]. The variations of the amplitude and the phase due to reflection, refraction and cross polarization are counted for the two polarizations by $(E_\perp, E_k)$ and $(\Phi_\perp, \Phi_k)$ respectively. They are calculated by the coordinate transformation matrix between two successive interaction points and the Fresnel formulas at the interaction point. $\Phi_P$ and $\Phi_F$, independent of polarization, are respectively the phase shifts due to the optical path difference and the focal lines [1, 16, 17].

After the amplitudes and the phases of all emergent rays are calculated with VCRM3D, we need to calculate the total electric field in order to get the scattered intensity in the given directions. Here, we are facing two difficulties. The first is that the directions of emergent rays $(\theta_i, \psi_i)$ (i being the index of emergent rays) are discrete and irregularly distributed, i.e. the amplitudes and the phases for both polarizations of all emergent rays are real functions in the form $f(\theta_i, \psi_i)$. So the interpolation is indispensable in order to calculate the complex amplitude in a given direction $(\theta, \psi)$. In VCRM2D and VCRM2D+, the interpolation is relatively easier because it is one-dimensional. For 3D scattering, the interpolation is two-dimensional. The second is that the emergent rays of the same order or different orders may be folded one or several times in certain regions, where the rays interfere with one another. Therefore, $f(\theta, \psi)$ is a multi-valued function. Even some standard 2D interpolation algorithms exist, such as that based on Delaunay triangulation in SciPy of Python, they are limited to single-valued functions. One needs to divide the calculated data $f(\theta_i, \psi_i)$ into single-valued zones before using those algorithms, which is also a difficult task. To address these problems, we have developed a line-by-line triangulation interpolation algorithm as follows.

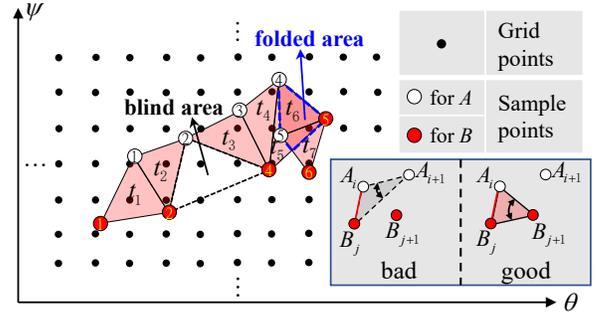

**Fig. 2.** Triangulation of the band formed by the emergent rays of two adjacent lines of incident rays.

The scattering space $\theta = [-180°, 180°]$ and $\psi = [-90°, 90°]$ is uniformly discretized into grid points as shown in Fig. 2. Each grid point (black dot) represents a scattering direction in 3D space at which the total electric field is to be calculated. Consider now two adjacent lines of incident rays, say $L_1$ and $L_2$ (Fig. 1), whose emergent rays are numbered according to the incident rays and shown in Fig. 2 respectively with white circles and red dots, called sets A and B. They are irregularly distributed in terms of scattering direction $(\theta_i, \psi_i)$. The band between the sets A and B is meshed incrementally with triangles $t_1, t_2, ...,$ as illustrated in Fig. 2. That means in generating a new triangle, the candidate vertex for the front edge $A_iB_j$ (i, j = 1, 2, ...) is always chosen as $A_{i+1}$ or $B_{j+1}$, and the successful one opens a larger angle with $A_iB_j$ as shown in the inset of Fig. 2. By this way, the long and thin triangles, which may deteriorate the interpolation accuracy, can be avoided to the maximum extent. Some points (e.g. ray 3 in set B) may be missing due to total internal reflection. Consequently, a blind area is formed and should be excluded in the triangulation. This triangulation procedure is repeated for the next adjacent lines of incident rays $L_2$ and $L_3$, then $L_3$ and $L_4$, and so on, until all emergent rays are dealt with. Then, inside each triangle, linear interpolation is carried out separately for the amplitude and the phase at each grid point being enclosed, according to the amplitudes and the phases of the three sample points (vertexes) [16]. The triangles meshed with those folded rays naturally overlap with each other as $t_5$, $t_6$ and $t_7$ do in Fig. 2. At $(\theta, \psi)$, assume the amplitude and the phase of polarization X ($\perp$ or k) interpolated by the N-th (N = 1, 2...) overlapping triangle are $E_{X,N}$ and $\Phi_{X,N}$, respectively. The sum of complex amplitudes accounts for the interference effect: $\tilde{E}_X(\theta, \psi) = \sum_{N=1} E_{X,N} \exp(i\Phi_{X,N})$. The intensity in the scattering direction $(\theta, \psi)$ is then obtained by: $I(\theta, \psi) = |\tilde{E}_\perp(\theta, \psi)|^2 + |\tilde{E}_k(\theta, \psi)|^2$.

Increasing the number of incident rays leads to smaller triangles and thereby improves the precision of interpolation. However, it is at the cost of computation time and memory. Two criteria are used to determine the number of incident rays. The first is based on the angle $\beta$ between the normal vectors of adjacent triangles (Fig. 3). The condition $\beta \leq 1°$ must be satisfied by no less than 98% of the triangles, with a tolerance of 2% reserved for the bad triangles generated when the amplitude or the phase experiences a sudden change, near the caustics for instance. The second is that the interpolation point number per degree should be greater than 2 times the angular frequency $f_{ripple}$ (in 1/degree) of the fine fringes [5] according to the Nyquist–Shannon sampling theorem.

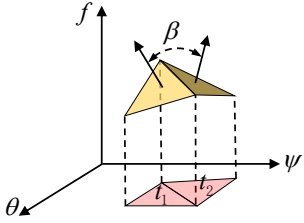

**Fig. 3.** Definition of angle $\beta$ between the normal vectors of two adjacent triangles, f represents the amplitude or the phase.

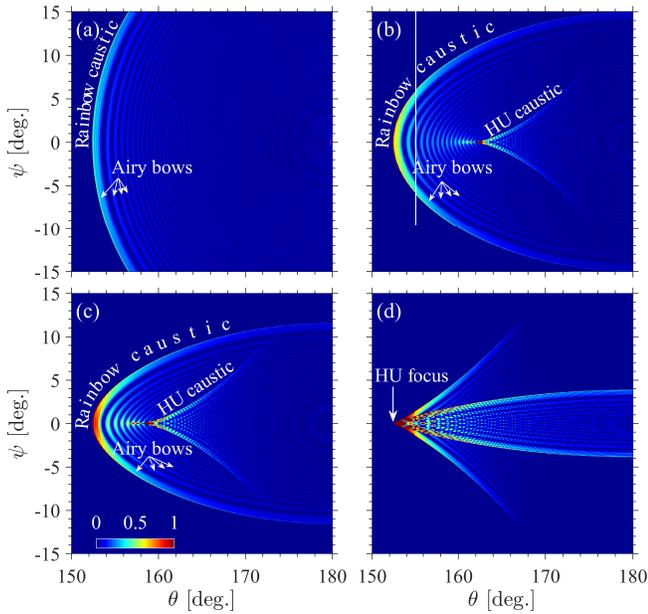

**Fig. 4.** Simulated scattering patterns near the primary rainbow of oblate drops (m = 1.4505, a = 143.31 $\mu$m) with different vertical semi-axis c illuminated by a plane wave $\lambda$ = 532.13 nm polarized along z axis: (a) c = a, (b) c = 130.78 $\mu$m, (c) c = 127.66 $\mu$m and (d) c = 119.87 $\mu$m. Color coding: intensity.

Fig. 4 examples the GRPs simulated for an oblate drop of Di-Ethyl-Hexyl-Sebacat in air [5] with different aspect ratios c/a. The rainbow of a spherical drop (a) is in circle form as predicted by Mie theory. It is deformed and becomes complex as the aspect ratio deviates from unity. The HU caustic [2, 6, 9] appears when the aspect ratio is not far from unity (c/a = 0.9126 and 0.8908 in (b) and (c) respectively). If the aspect ratio decreases further (c/a = 0.8364 in (d)) the HU focus comes out. These phenomena correspond well to those reported in the literature [2, 4–7, 9]. The benefit of VCRM3D, compared to simple ray-tracing model [4, 6, 7, 9] and its previous 2D and 2D+ implementations [1, 15], is now clear. It permits to calculate the total intensity of all the 3D rays arriving in the same direction, i.e. their interference and thus to predict all the coarse and fine structures of GRPs.

To prove the capacity of our method, we examine now in detail the simulated GRPs. Fig. 5 (a) shows the skeleton (HU caustic) simulated with our method, which is exactly the same as that of pure GO. Fig. 5 (b) is a zoom of Fig. 4 (c) to be compared with Fig. 5 (c) the experiment pattern (extracted from Fig. 2 (b) in [5]). In the two-ray zone, two rays of the same order p = 2 emerge in a given direction, indicating the wavefront is folded up once. The two rays interfere forming the Airy bows. In the four-ray zone, the wavefront is folded up twice, producing a lattice pattern, whose frontier with the two-ray zone is the HU caustic (dashed line in Fig. 5 (a)). The ripple fringes discernable on the Airy bows are due to the interference of the rays of orders p = 0 and p = 2. The distance between the fringes and their inclination are very sensible to the size and the aspect ratio of the drop, so useful for particle characterization. Nevertheless, we observe an abrupt variation of intensity along the frontier between the two-ray and four-ray regions, i.e. HU caustic, in the simulated pattern. This discontinuity is due to the intrinsic deficiency of ray model and may be remedied by taking into account the diffraction effect near the caustics. This is another story and beyond the scope of this letter.

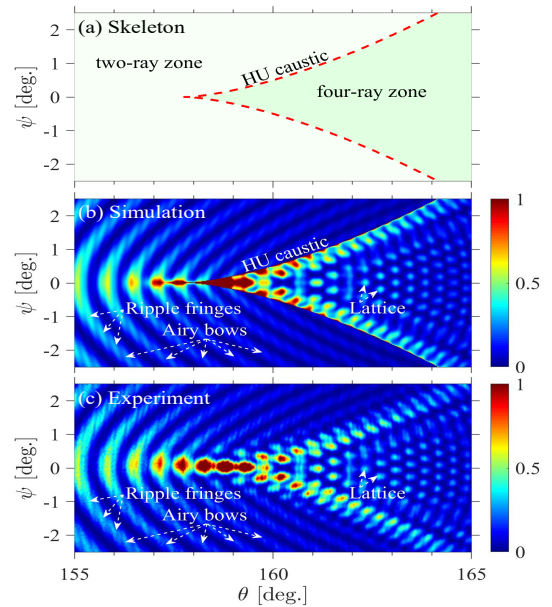

**Fig. 5.** Comparison of GRPs: (a) Caustic structure, (b) Zoom of Fig. 4(c) illustrating details of the simulated GRP, and (c) Experimental GRP (Fig. 2 (b) in [5]).

The computation time for the simulation of a full view of GRP in Fig. 4 takes about 30 minutes on a desktop computer (AMD Ryzen 7 1700, Eight-Core Processor, RAM 16 GB), of which about 20 minutes are spent on the calculation of the amplitudes and the phases of all the emergent rays ($p \leq 2$) of 4 million incident rays, and 10 minutes on the triangulation and the interpolation for 1500 × 1500 grid points in the angle ranges $\theta$ = [150°, 180°] and $\psi$ = [−15°, 15°], i.e. an angular resolution of 0.02°.

The equivalent size parameters $2\pi r/\lambda$ of the drops in our simulation are over 1500, with r being the radius of the equivalent-volume sphere. There is no rigorous light scattering model yet for a direct comparison. To evaluate the precision of our simulation, we make a quantitative comparison of the simulated result with the experimental one and that of the VCRM2D+ in Fig. 6, for the scattered intensity in the horizontal plane ($\psi = 0°$). We note firstly that the three curves are in very good agreement in the two-ray zone between about 152.6° and 157.2°. The discrepancy at borders of this zone, near and below rainbow angle and in the vicinity of HU cusp, is evident as pointed out above. In the four-ray zone $\theta > 158.5°$ the intensity simulated with VCRM3D is in very good agreement with the experimental result in the range from 158.5° to 162°, but the discrepancy becomes visible after 162°. This discrepancy is probably caused by a slight deformation of the drop in the experiment and/or the little difference between the real drop size and that used in the simulation. These influences will be investigated below. The intensity calculated with VCRM2D+ (the model and the software available in [1, 18]) does not agree with the experimental result in the four-ray zone. This is due to the fact that VCRM2D+ takes into account only the rays in the equatorial plane, but without the skew rays [2, 6] which illuminate the drop above and below the equatorial plane but emerge in the direction parallel to the equatorial plane.

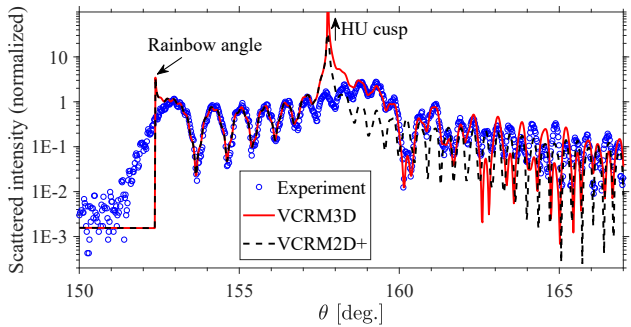

**Fig. 6.** Comparison of the scattering diagrams of the GRP in horizontal plane obtained by experiment [5] and by simulations with VCRM3D and VCRM2D+ [1, 18]. The parameters are the same as in Fig. 4(c).

We investigate now the sensitivity of the scattering pattern in the four-ray zone to the aspect ratio of the oblate drop. The scattered intensities in the equatorial plane of four oblate drops with very little difference in the vertical semi-axis c are shown in Fig. 7. The drop's semi-axis a is taken to be constant 143.31 $\mu$m. For the rays in the equatorial plane, a small deformation of the drop in the vertical direction brings only tiny variations to the divergence factor, therefore the amplitude. That is why the scattered intensity in the two-ray zone is not so sensitive to the deformation of the drop. But in the four-ray zone, the contributions of the skew rays make the scattering pattern much more sensitive to the aspect ratio of the drop. This is because the value of c affects not only the divergence factor, but also the incident and the emergent positions of skew rays, hence the phase and polarization state. The capability to predict the light scattering in the four-ray zone, where the scattering pattern shows high sensitivity to the drop's aspect ratio, may improve the precision in the characterization of the drop's shape.

In summary, a new achievement for 3D scattering of a large non-spherical particle is realized in the framework of VCRM with the line-by-line triangulation interpolation algorithm. The qualitative and quantitative comparisons of the simulated and experimental GRPs of oblate drops demonstrate clearly that our method allows, for the first time, taking into account all the rays in 3D space and predicting in fine the total scattering patterns correctly in all space except in the vicinity of caustics. The improvement for the scattered intensity in the immediate vicinity of caustics is in progress. Though this letter is focused on the oblate drops, the developed method can be applied directly to large drops or bubbles of other shapes with smooth surfaces [19–21]. The applications of the VCRM3D in other domains, such as freeform optics, characterization of liquid ligaments, are to be investigated.

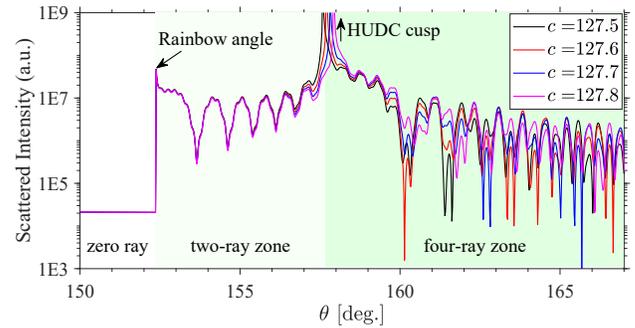

**Fig. 7.** Sensitivity of the scattered intensity in the equatorial plane to the vertical semi-axis c of an oblate drop with a = 143.31 $\mu$m.

**Funding.** This work was supported by the Fundamental Research Funds for the Central Universities of China [grant number XJS210507], the 111 Project of China [grant number B17035] and the French National Research Agency [grant number ANR-13-BS090008-01, BS090008-02 (AMO-COPS)].

**Disclosures.** The authors declare no conflicts of interest.

**Data availability.** Data underlying the results presented in this paper are not publicly available at this time but may be obtained from the authors upon reasonable request.